\documentclass[12pt]{article}

\usepackage{latexsym,graphicx,graphics,eufrak}
\usepackage{amsmath,epsfig,abstract,authblk}
\usepackage{appendix}
\usepackage[margin=1in]{geometry} 
\newcommand{\abs}[1]{\left| #1 \right|}

\newcommand{\bb}[1]{ \mbox{\boldmath$ #1$}}

\newcommand{\Eq}[1]{Eq.~(\ref{#1})}

\begin{document}
\title{Clustering in particle chains - summation techniques for the periodic Green's function}
\author{Yarden Mazor\footnote{yardenm2@mail.tau.ac.il}, Yakir Hadad, Ben Z. Steinberg}
\affil{School of Electrical Engineering, Tel Aviv University, Ramat-Aviv, Tel-Aviv 69978  Israel}
%\email{yardenm2@mail.tau.ac.il}
\maketitle
\begin{abstract}
1D lattice summations of the 3D Green's function are needed in many applications such as photonic crystals, antenna arrays, and so on. Such summations are usually divided into two cases, depending on the location of the observer: Out of the summation axis, or on the summation axis. Here, as a service for the community, we present and summarize the summation formulas for both cases. On the summation axis, we use polylogarithmic functions to express the summation, and Away from the summation axis we use Poisson summation (equivalent to the expansion of the field to cylindrical harmonics)
\end{abstract}

This text is not meant to be a comprehensive overview of the literature in this topic. We have included several references to selected works that incorporate parts of this overview, or other related methods. If someone feels that we have missed or did not credit his work in related matters, please do not hesitate to approach us, and we would gladly revise the bibliography. 
\section{Introduction}
\begin{figure}[htbp]
\centering
\includegraphics[scale=0.75]{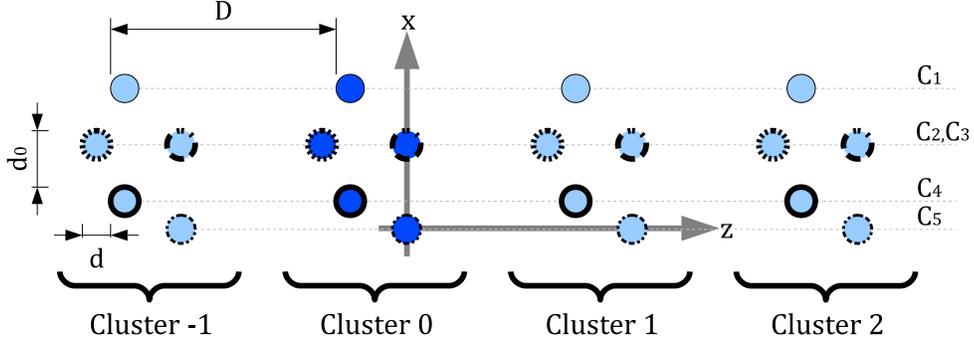}
\caption{General configuration of a clustered chain}
\label{fig1}
\end{figure}

The most general form of a cluster in a particle chain is shown in Fig.~\ref{fig1}.
The theoretical "Construction" process of such a structure may be visualised as taking a $N$ basic chains composed of particles with polarizability $\alpha_n, (n=1,...,N)$ and inter-particle distance $D$, and placing them in some general configuration, so all the chains have colinear chain axes (in this case visualised by the parallel dashed gray lines). If we assume the particles are small enough with respect to the inter-particle distance and the wavelength we may use the Discrete Dipole Approximation. In addition we assume that if as small particle with polarizability $\bb{\alpha}$ is subject to an electric field whose local value in the absence of the particle is $\bb{E^L}$ it will respond by forming an electric dipole moment $\bb{p}=\bb{\alpha E^{L}}$. if we mark the dipoles induced in each particle by $\bb{p_{nm}}$ where $n$ enumerates the particles inside the cluster $(n=[1,...,N])$, and $m$ will be the cluster index. 

It is worth noting before we continue, that particular cases of the formulation presented in this summary can be found in many works. \cite{AluParallel} treated simple cases of 2 parallel chains, \cite{LongCh} dealt with general longitodinal particle clusters. Cases involving various summation techniques in 2D can be found in \cite{TretyakovBook},\cite{Cappolino_PRE_2011},\cite{Vitaly}. 

We will first formulate the most general case, and then continue by treating each of the basic "building blocks" separately. 
\section{Formulation}
 The local electric field value for each of the particles may then be written in the most general form as
	\begin{equation}
	\bb{E^L}=\sum_{n=1}^{N}\sum_{m=-\infty}^{\infty}\bb{G}\left(\bb{r_0},\bb{r_{nm}}\right)\bb{p_{nm}}
	\label{eq1}
	\end{equation}
where $\bb{G}$ is the dyadic Green's function in free space, $\bb{r_0}$ is the location of the particle under examination, and $\bb{r_{nm}}$ is the location of the $\bb{p_{nm}}$ dipole.
Since the structure is periodic with a period $D$, we may express each dipole along the chain as 
	\begin{equation}
	p_{nm}=p_{n0}e^{i\beta mD}
	\label{eq2}
	\end{equation}
Since the system is periodic, we may treat only the particles that reside in \emph{cluster 0} (noted in darker blue in Fig.~\ref{fig1}). If we mark the particle we are examining in \emph{cluster 0} as the $n'th$ particle, taking the summation in \Eq{eq1} and multiplying from the left by $\bb{\alpha_{n'}^{-1}\alpha_{n'}}$ will result in the equation
	\begin{equation}
	\bb{\alpha_{n'}^{-1}}\bb{p_{n'0}}= \sum_{n=1}^{N}\sum_{m=-\infty}^{\infty}\bb{G}\left(\bb{r_0},\bb{r_{nm}}\right)\bb{p_{n0}}e^{i\beta mD}
	\label{eq3}
	\end{equation}
or in a more intuitive form
	\begin{equation}
	\bb{\alpha_{n'}^{-1}}\bb{p_{n'0}}= \sum_{n=1}^{N}\left(\sum_{m=-\infty}^{\infty}\bb{G}\left(\bb{r_0},\bb{r_{nm}}\right)e^{i\beta mD}\right)\bb{p_{n0}}
	\label{eq4}
	\end{equation}
A more accurate way to write this system of equations is by separating $n=n'$ from the rest of the possible values of $n$ which gives the form
	\begin{equation}
	\left(\sum_{\substack{m=-\infty \\ m\neq 0}}^{\infty}\bb{\bar{G}}\left(\bb{r_0},\bb{r_{n'm}}\right)e^{i\beta mD}-\bb{\alpha_{n'}^{-1}}\right)\bb{p_{n'0}}+ 
	\sum\limits_{\substack{n=1 \\ n\neq n'}}^{N}\left(\sum_{m=-\infty}^{\infty}\bb{\bar{G}}\left(\bb{r_0},\bb{r_{nm}}\right)e^{i\beta mD}\right)\bb{p_{n0}}=0
	\label{eq5}
	\end{equation}
Where $\bb{\bar{G}}=\frac{6\pi\epsilon_0}{k^3}\bb{G}$. Each value of n' defines a certain particle to be examined, and in fact defines a certain \emph{Base Chain} we are treating. The different \emph{Base Chains} from which we construct a more general \emph{Clustered Chain} are noted in Fig.~\ref{fig1} on the right side as C1,C2,C3,C4,C5. The last equation defines a $3N\times 3N$ matrix equation where each value selected for n' essentially defines the lines $3n'-2, 3n'-1, 3n'$ of it. The matrix representing the entire system may also be described as $N\times N$ block matrix $\mathfrak{M}$, where each block is $3\times 3$  and defines the interaction between a certain\emph{base chain} to another. Using the block-matrix notation the equation may be written as $\mathfrak{M}\cdot\bb{p}=0$ or 
	\begin{equation}
	\begin{pmatrix}
	\bb{M_{11}} & \bb{M_{12}} & \cdots & \bb{M_{1N}} \\
	\bb{M_{21}} & \ddots &   & \vdots \\
	\vdots & & & \vdots \\
	\bb{M_{N1}} & \cdots & \cdots & \bb{M_{NN}} 
	\end{pmatrix}
	\begin{pmatrix}
	\bb{p_{10}} \\
	\bb{p_{20}} \\
	\vdots \\
	\bb{p_{N0}}
	\end{pmatrix}
	=0
	\label{eq6}
	\end{equation}
The value of $M_{n',q'}$ depends mostly on the geometrical positioning of the \emph{base chain} p' in relation to the \emph{base chain} of reference n'. This positioning is defined by 2 parameters  - the longitudinal shift $d$ and the transverse shift $d_{0}$. These parameters are illustrated in Fig.~\ref{fig1} for C2 and C4.
A non-trivial solution for this system exists only if
	\begin{equation}
	det(\mathfrak{M})=0
	\label{eq7}
	\end{equation}

\section{The diagonal terms of $\mathfrak{M}$}
The diagonal terms arise from the left brackets in \Eq{eq5} and are well-known from work done on simple particle chains (not clustered), Therefore the diagonal block $M_{n'n'}$ may be written as
	\begin{equation}
	M_{n'n'}=
	\begin{pmatrix}
	T & 0 & 0 \\
	0 & T & 0 \\
	0 & 0 & L 
	\end{pmatrix}
	\label{eq8}
	\end{equation}
where
	\begin{equation}
	\begin{array}{rcl}
	T&=&\frac{3}{2}\left[ \frac{1}{kD}f_{1}(kD,\beta D)+ \frac{i}{(kD)^2}f_{2}(kD,\beta D)- \frac{1}{(kD)^3}f_{3}(kD,\beta D) \right] - \bar{\alpha}_{n'}^{-1} \\
	\\
	L&=&3\left[ - \frac{i}{(kD)^2}f_{2}(kD,\beta D)+ \frac{1}{(kD)^3}f_{3}(kD,\beta D) \right] - \bar{\alpha}_{n'}^{-1} \\
	\label{eq9}
	\end{array}
	\end{equation}
where
\begin{equation}
f_s(x,y)=Li_s[e^{i(x+y)}]+Li_s[e^{i(x-y)}]
\label{eq9b}
\end{equation}
This block represents the interaction between a certain \emph{base chain} with itself.

\section{Terms of $\mathfrak{M}$ which represent $d_{0}=0$}
\begin{figure}[htbp]
	\centering
	\includegraphics[scale=0.75]{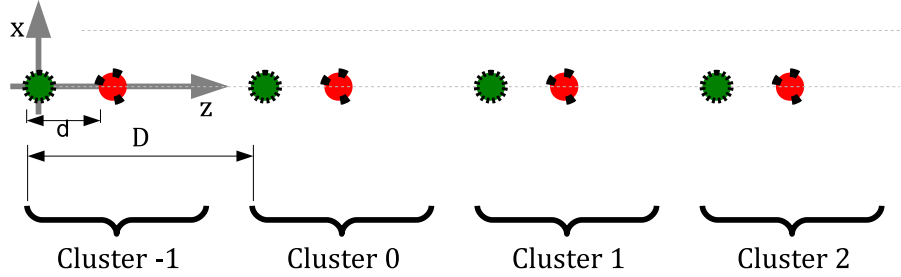}
	\caption{General configuration of a clustered linear chain}
	\label{fig2}
	\end{figure}

These terms represent the interaction of 2 chains that are completely co-linear with each other (Share a common chain axis). In Fig.~\ref{fig1} the \emph{base chains} that posses such property are C2,C3. This case is "isolated" in Fig.~\ref{fig2}. 
 The $\bb{M}_{n',q'}$ term corresponding to this case has the form
	\begin{equation}
	\bb{M}_{n',q'}= \frac{6\pi}{k^3} \sum_m\bb{A}(mD+d)e^{i\beta mD}
	\label{eq10}
	\end{equation}
Where
	\begin{equation}
	\bb{A}(z)=\frac{e^{ik\abs{z}}}{4\pi\abs{z}}
	\left[k^2\bb{A}_1+\left(\frac{1}{z^2}-\frac{ik}{\abs{z}}\right)\bb{A}_2\right]\label{eq11}
	\end{equation}
Is the dyadic Green's function simplified for the case of particles residing all on the Z-axis and $\bb{A}_1=\mbox{diag}(1,1,0),\bb{A}_2=\mbox{diag}(-1,-1,2)$. In order to give a simple solution to the summation given in \Eq{eq11} we are required to assume that the given inter-particle distance $d$ is a rational fraction of the chain period $D$. This requires that the distances $d,D$ are both integral multiplications of some basic distance $\delta$
meaning
	\begin{equation}
	D=L\delta,\;\; d=\ell\delta
	\end{equation}
Under this assumption the summation in \Eq{eq11} may be rewritten as
	\begin{equation}
	\bb{M}_{n',q'}=h_1[\ell]\bb{A}_1-ih_2[\ell]\bb{A}_2+h_3[\ell]\bb{A}_2\label{eq12}
	\end{equation}
where 
	\begin{equation}
	h_s[\ell]=\frac{3}{2}\frac{1}{(k\delta)^s}\sum_m \frac{e^{ikd\abs{mL+\ell}+i\beta \delta mL}}{\abs{mL+\ell}^s}.\label{eq13}
	\end{equation}

\subsection{Evaluation of $h_s[\ell]$}
For $1\le\ell\le L-1$, $h_s[\ell]$ can be re-written as
	\begin{equation}
	h_s[\ell] = \frac{3}{2}\frac{e^{-i\beta\delta\ell}}{(k\delta)^s}\sum_{m=0}^\infty
	\frac{\left[e^{i(k+\beta)\delta}\right]^{mL+\ell}}{(mL+\ell)^s}\, +  \frac{3}{2}\frac{e^{-i\beta\ell \delta}}{(k\delta)^s}\sum_{m=1}^\infty
	\frac{\left[e^{i(k-\beta)\delta}\right]^{mL-\ell}}{(mL-\ell)^s}.\label{eq14}
	\end{equation}
We concentrate on the first sum above. It has the form
	\begin{equation}
	\sigma=\sum_{m=0}^\infty\frac{(e^{ix})^{mL+\ell}}{(mL+\ell)^s}.\label{eq15}
	\end{equation}
This sum can be re-written as
	\begin{equation}
	\sigma=\sum_{n=1}^\infty \frac{(e^{ix})^n}{n^s}\cdot a_n(\ell)\label{eq16}
	\end{equation}
where $a_n(\ell)$ is an \emph{auxiliary periodic sequence} of period $N$, satisfying
	\begin{equation}
	a_n(\ell)=\left\{
	\begin{array}{ll}
	0,\quad & n=1,\ldots , L,\,\, n\ne\ell\\
	1, &n=\ell\end{array}\right. .\label{eq17}
	\end{equation}
Clearly, the series periodicity implies the recurrence relation $a_{n+L}=a_n$, whose characteristic polynomial $p(\lambda)=\lambda^L-1$ has $N$ distinct roots
	\begin{equation}
	\lambda^L-1=0\,\Rightarrow\, \left\{\lambda_r\right\}_{r=0}^{L-1}=e^{i2\pi r/L},\label{eq18}
	\end{equation}
hence the \emph{infinite} sequence $a_n(\ell),\,\forall\, n$ can be generated by the \emph{finite} sum
	\begin{equation}
	a_n(\ell)=\sum_{r=0}^{L-1} C_r(\ell)\lambda_r^n=\sum_{r=0}^{L-1} C_r(\ell) e^{i2\pi r n/L}.\label{eq19}
	\end{equation}
The coefficients $C_r(\ell)$ can be determined using the $L$-initial conditions of the recurrence relation [given by \Eq{eq17}] in \Eq{eq19}. The result is the $L\times L$ \emph{Vandermonde} matrix equation $\bb{a}(\ell)=\bb{\Lambda}\bb{C}(\ell)$. Here $\bb{a}(\ell)$ is a vector of $L$ entries, whose elements are given by \Eq{eq15}, $\bb{C}(\ell)$ is the vector of unknown coefficients, and $\bb{\Lambda}$ is a Vandermonde matrix, whose $n,r$ entry is $\Lambda_{nr}=e^{i2\pi rn/L}$. Due to its specific structure, $\bb{\Lambda}$ is also a unitary transformation from the Euclidean basis $a_n(\ell)$ (roam with $\ell$) to a Fourier basis. Its inverse is its adjoint (normalize first). Hence
	\begin{eqnarray}
	C_r(\ell)&=& L^{-1}e^{-i2\pi r\ell/L}\,\Rightarrow\nonumber\\
	a_n(\ell)&=& \frac{1}{L}\sum_{r=0}^{L-1}e^{i2\pi r (n-\ell)/L},\,\forall n.\label{eq20}
	\end{eqnarray}
Substituting this result into \Eq{eq16} and exchanging the order of summation, we can express $\sigma$ as a finite sum of Polylogarithms
	\begin{equation}
	\sigma=L^{-1}\sum_{r=0}^{L-1}e^{-i2\pi r\ell/L}Li_s(e^{ix+i2\pi r/L}).\label{eq21}
	\end{equation}
Likewise, we may repeat essentially the same procedure for the second sum in \Eq{eq14} (note the lower summation bound; shift the index by 1, and at the end change $r\mapsto L-r'$). The final result for $h_s[\ell],\, 1\le\ell\le L-1,$ is
	\begin{eqnarray}
	h_s[\ell]&=& \frac{3}{2L}\frac{e^{-i\beta \delta\ell}}{(k\delta)^s}\sum_{r=0}^{L-1} e^{-i2\pi r \ell/L} f_s(k\delta,\beta \delta + 2\pi r/L)\label{eq22}
	\end{eqnarray}
and $f_s$ are given in \Eq{eq9b}. Though developed for the case of $1\le\ell\le L-1$ the expression given in \Eq{eq22} is in fact valid for all values of $\ell$ that satisfy $1\le|\ell|\le L-1$.
As a last remark for this section we mention that even though these formulas were developed here from "scratch", one could find many similarities to relations from signal processing where conversions of sampling rate take place.

\section{Terms of $\mathfrak{M}$ which represent $d_{0}\neq 0$}
\begin{figure}[htbp]
	\centering
	\includegraphics[scale=0.75]{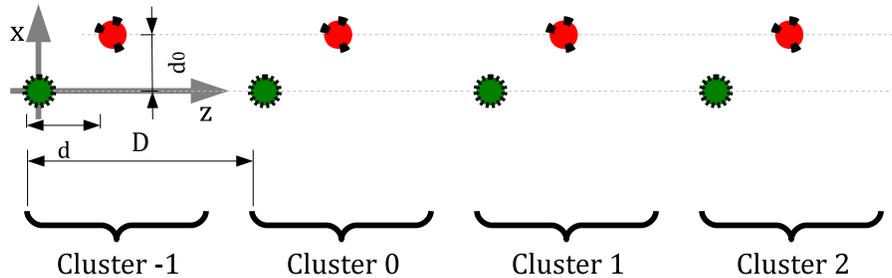}
	\caption{General configuration of a transversely clustered chain}
	\label{fig3}
	\end{figure}

This case corresponds to interactions between chains as shown in Fig.~\ref{fig3}.
This case presents us with a different challenge since the transverse shift between the \emph{base chains} couples different polarizations such as X and Z polarizations. In order to properly treat this type of interactions we will write the corresponding term $\bb{M}_{n',q'}$ as
	\begin{equation}
	\bb{M}_{n',q'}=
	\begin{pmatrix}
	{M}_{n',q'}^{xx} & {M}_{n',q'}^{xy} & {M}_{n',q'}^{xz} \\
	{M}_{n',q'}^{yx} & {M}_{n',q'}^{yy} & {M}_{n',q'}^{yz} \\
	{M}_{n',q'}^{zx} & {M}_{n',q'}^{zy} & {M}_{n',q'}^{zz} 
	\end{pmatrix}
	\label{eq23}
	\end{equation}
Since we are discussing only 2D planar clustering of particles (the XZ plane) $M_{n',q'}^{xy}={M}_{n',q'}^{yx}={M}_{n',q'}^{yz}={M}_{n',q'}^{zy}=0$.
In addition this block has to be symmetrical due to the symmetry in Green's dyad and therefore ${M}_{n',q'}^{xz}={M}_{n',q'}^{zx}$. This enables to write down the block as
	\begin{equation}
	\bb{M}_{n',q'}=
	\begin{pmatrix}
	{M}_{n',q'}^{xx} & 0 & {M}_{n',q'}^{xz} \\
	0 & {M}_{n',q'}^{yy} & 0 \\
	{M}_{n',q'}^{xz} & 0 & {M}_{n',q'}^{zz} 
	\end{pmatrix}
	\label{eq25}
	\end{equation}
In the following analysis we will use the following 
	\begin{equation}
	R_m=\sqrt{d_0^2+(mD-d)^2}, \bb{r}_m=(d_0,0,mD)
	\label{eq26}
	\end{equation}
The main tool we are going to use to evaluate the terms in the $\bb{M}_{n',q'}$ block is the Poisson summation formula. this will allow us to "convert" the algebraic summations  to summations over a series of Hankel functions. since for large arguments we may obtain exponential-like decay in Hankel functions, this allows us to sum a finite number of elements in the series and still obtain a very good approximation for the terms in the matrix.
\subsection{Evaluation of ${M}_{n',q'}^{xx}$}\label{xx}
The value of the x component of $\bb{E}^L$ that the red chain creates at the origin
	\begin{equation}
	\bb{E}_{x}^{L}(\bb{r}=(0,0,0))=\sum_{m=-\infty}^{\infty}G_{xx}(0,\bb{r}_m)p_{0x}e^{i\beta mD}
	\label{eq27}
	\end{equation}
where $p_{0x}e^{i\beta mD}$ are the X components of the dipole moments induced on the m'th particle in the red chain.
substituting $G_{xx}$ we may obtain an explicit form for ${M}_{n',q'}^{xx}$
	\begin{equation}
	{M}_{n',q'}^{xx}=\frac{3}{2}\sum_{m=-\infty}^{\infty}e^{ikR_m}\left[\frac{1}{kR_m}-\frac{(kd_0)^2}{(kR_m)^3}-\frac{1}{(kR_m)^3}+i\frac{1}{(kR_m)^2}+3\frac{(kd_0)^2}
	{(kR_m)^5}-3i\frac{(kd_0)^2}{(kR_m)^4}\right]e^{i\beta mD}
	\label{eq28}
	\end{equation}
For easier evaluation this may also be written as
	\begin{equation}
	{M}_{n',q'}^{xx}=\frac{3}{2}\sum_{m=-\infty}^{\infty}\left[\frac{e^{ikR_m}}{kR_m}+\frac{\partial^2}{\partial (kd_0)^2}\frac{e^{ikR_m}}{kR_m}\right]e^{i\beta mD}
	\label{eq29}
	\end{equation}
We start by evaluating the summation over the left term in \Eq{eq29}. Substituting $R_m$ we get
	\begin{equation}
	\frac{3}{2}\sum_{m=-\infty}^{\infty}\frac{e^{ikR_m}}{kR_m}=\frac{3}{2}\sum_{m=-\infty}^{\infty}\frac{e^{ik\sqrt{d_0^2+(mD-d)^2}}}{k\sqrt{d_0^2+(mD-d)^2}}
	\label{eq30}
	\end{equation}
To evaluate this sum we use Poisson summation formula
	\begin{equation}
	\sum_{m=-\infty}^{\infty}f(m)=\sum_{n=-\infty}^{\infty}\int_{-\infty}^{\infty}f(x')e^{-2\pi inx'}dx'
	\label{eq31}
	\end{equation}
Substituting our expressions into \Eq{eq31} we obtain
	\begin{equation}
	\frac{3}{2}\sum_{m=-\infty}^{\infty}\frac{e^{ik\sqrt{d_0^2+(mD-d)^2}}}{k\sqrt{d_0^2+(mD-d)^2}}e^{i\beta mD}=\frac{3}
	{2}\sum_{n=-\infty}^{\infty}\int_{-\infty}^{\infty}\frac{e^{ik\sqrt{d_0^2+(x'D-d)^2}}}{k\sqrt{d_0^2+(x'D-d)^2}}e^{i\beta x'D}e^{-2\pi inx'}dx'
	\label{eq32}
	\end{equation}
Evaluation of such an integral is possible using integral tables
	\begin{equation}
	\int_{-\infty}^{\infty}\frac{e^{ik\sqrt{d_0^2+(x'D-d)^2}}}{k\sqrt{d_0^2+(x'D-d)^2}}e^{i\beta x'D}e^{-2\pi inx'}=
	-\frac{e^{\left(\frac{\beta}{k}-\frac{2\pi n}{kD}\right)kd}}{kD}\frac{\pi}{i}H_{0}^{(1)}\left[kd_0\sqrt{1-\left(\frac{2\pi n}{kD}-\frac{\beta}{k}\right)^2}\right]
	\label{eq33}
	\end{equation}
we may define the normalized parameters for easier notation
	\begin{equation}
	\bar{d}=kd,\;\;\bar{d}_0=kd_0,\;\;\bar{D}=kD,\;\;\tilde{\beta}_{n}=\frac{2\pi n}{\bar{D}}-\frac{\beta}{k}
	\label{eq34}
	\end{equation}
and finally obtain
	\begin{equation}
	\frac{3}{2}\sum_{m=-\infty}^{\infty}\frac{e^{ikR_m}}{kR_m}=-\frac{3}{2}\sum_{n=-\infty}^{\infty}\frac{e^{-i\tilde{\beta}_{n}\bar{d}}}{\bar{D}}\frac{\pi}
	{i}H_{0}^{(1)}\left[\bar{d}_0\sqrt{1-\tilde{\beta}_{n}^2}\right]
	\label{eq35}
	\end{equation}
We may treat the summation over the right term in \Eq{eq29} using the same method and applying the needed derivatives we obtain
	\begin{equation}
	\begin{split}
	\frac{3}{2}\sum_{m=-\infty}^{\infty}\frac{\partial^2}{\partial (kd_0)^2}\frac{e^{ikR_m}}{kR_m}=\frac{3}{2}\sum_{n=-\infty}^{\infty}\frac{e^{-i\tilde{\beta}_{n}\bar{d}}}
	{\bar{D}}\frac{\pi}{i}\left(1-\tilde{\beta}_{n}^2\right)\cdot \\
	\cdot\frac{1}{2}\left(1-\tilde{\beta}_{n}^2\right)\left\{H_{0}^{(1)}\left[\bar{d}_0\sqrt{1-\tilde{\beta}_{n}^2}\right]-
	H_{2}^{(1)}\left[\bar{d}_0\sqrt{1-\tilde{\beta}_{n}^2}\right]\right\}
	\label{eq36}
	\end{split}
	\end{equation}
Adding \Eq{eq35} and \Eq{eq36} we obtain the final expression
	\begin{equation}
	{M}_{n',q'}^{xx}=\frac{3}{2}\sum_{n=-\infty}^{\infty}\frac{e^{-i\tilde{\beta}_{n}\bar{d}}}{\bar{D}}\frac{\pi}{i}
	\cdot\left\{-\frac{1}{2}\left(\tilde{\beta}_{n}^2+1\right)H_{0}^{(1)}\left[\bar{d}_0\sqrt{1-\tilde{\beta}_{n}^2}\right]+
	\frac{1}{2}\left(\tilde{\beta}_{n}^2-1\right)H_{2}^{(1)}\left[\bar{d}_0\sqrt{1-\tilde{\beta}_{n}^2}\right]\right\}
	\label{eq37}
	\end{equation}
Where $H^{(1)}_{n}$ are Hankel functions of the first kind, of order $n$. We will repeat a very similar process for the other $M$ terms in the block.

\subsection{Evaluation of ${M}_{n',q'}^{yy}$}\label{yy}
The evaluation of ${M}_{n',q'}^{yy}$ is somewhat less tedious since y polarization is not coupled with other polarizations. The value of the y component of $\bb{E}^L$ that the red chain creates at the origin due to the y components of the dipoles in the red chain
	\begin{equation}
	\bb{E}_{y}^{L}(\bb{r}=(0,0,0))=\sum_{m=-\infty}^{\infty}G_{yy}(0,\bb{r}_m)p_{0y}e^{i\beta mD}
	\label{eq38}
	\end{equation}
where $p_{0y}e^{i\beta mD}$ are the y components of the dipole moments induced on the m'th particle in the red chain.
substituting $G_{yy}$ we may obtain an explicit form for ${M}_{n',q'}^{yy}$
	\begin{equation}
	{M}_{n',q'}^{yy}=\frac{3}{2}\sum_{m=-\infty}^{\infty}e^{ikR_m}\left[\frac{1}{kR_m}+\frac{i}{(kR_m)^2}-\frac{1}{(kR_m)^3}\right]e^{i\beta mD}
	\label{eq39}
	\end{equation}
This may also be written as 
	\begin{equation}
	{M}_{n',q'}^{yy}=\frac{3}{2}\sum_{m=-\infty}^{\infty}\left[\frac{e^{ikR_m}}{kR_m}+\frac{1}{\bar{d}_0}\frac{\partial}{\partial\bar{d}_0}\frac{e^{ikR_m}}{kR_m}\right]e^{i\beta mD}
	\label{eq40}
	\end{equation}
Summation over the left term gives the result presented in \Eq{eq35}. After performing the derivatives summation over the right term gives 
\begin{equation}
\frac{3}{2}\sum_{m=-\infty}^{\infty}\frac{1}{\bar{d}_0}\frac{\partial}{\partial\bar{d}_0}\frac{e^{ikR_m}}{kR_m}=
\frac{3}{2}\sum_{n=-\infty}^{\infty}\frac{e^{-i\tilde{\beta}_{n}\bar{d}}}{\bar{D}}\frac{\pi}{i}\frac{\sqrt{1-\tilde{\beta}_{n}^2}}{\bar{d}_0}
H_{1}^{(1)}\left[\bar{d}_0\sqrt{1-\tilde{\beta}_{n}^2}\right]
\label{eq41}
\end{equation}
Adding \Eq{eq35} and \Eq{eq41}
	\begin{equation}
	{M}_{n',q'}^{yy}=\frac{3}{2}\sum_{n=-\infty}^{\infty}\frac{e^{-i\tilde{\beta}_{n}\bar{d}}}{\bar{D}}\frac{\pi}{i}
	\cdot\left\{-H_{0}^{(1)}\left[\bar{d}_0\sqrt{1-\tilde{\beta}_{n}^2}\right]+
	\frac{\sqrt{1-\tilde{\beta}_{n}^2}}{\bar{d}_0}H_{1}^{(1)}\left[\bar{d}_0\sqrt{1-\tilde{\beta}_{n}^2}\right]\right\}
	\label{eq42}
	\end{equation}

\subsection{Evaluation of ${M}_{n',q'}^{zz}$}\label{zz}
The value of the z component of $\bb{E}^L$ that the red chain creates at the origin due to the z components of the dipoles in the red chain 
	\begin{equation}
	\bb{E}_{z}^{L}(\bb{r}=(0,0,0))=\sum_{m=-\infty}^{\infty}G_{zz}(0,\bb{r}_m)p_{0z}e^{i\beta mD}
	\label{eq43}
	\end{equation}
where $p_{0z}e^{i\beta mD}$ are the z components of the dipole moments induced on the m'th particle in the red chain.
substituting $G_{zz}$ we may obtain an explicit form for ${M}_{n',q'}^{zz}$
	\begin{equation}
	\begin{split}
	{M}_{n',q'}^{zz}=\frac{3}{2}\sum_{m=-\infty}^{\infty}e^{ikR_m}\left[\frac{1}{kR_m}-\frac{(\bar{d}-m\bar{D})^2}{(kR_m)^3}-\frac{1}{(kR_m)^3}+i\frac{1}{(kR_m)^2}\right.\\
	\left.+3\frac{(\bar{d}-m\bar{D})^2}{(kR_m)^5}-3i\frac{(\bar{d}-m\bar{D})^2}{(kR_m)^4}\right]e^{i\beta mD}
	\label{eq44}
	\end{split}
	\end{equation}
This may also be written as 
	\begin{equation}
	{M}_{n',q'}^{zz}=\frac{3}{2}\sum_{m=-\infty}^{\infty}\left[\frac{e^{ikR_m}}{kR_m}+\frac{1}{\bar{d}_0}\frac{\partial^2}{\partial(\bar{d}-m\bar{D})^2}\frac{e^{ikR_m}}{kR_m}\right]e^{i\beta mD}
	\label{eq45}
	\end{equation}
Again, Summation over the the left term is given by \Eq{eq35}. Using Fourier transform properties, we may write an expression for the summation over the right term
	\begin{equation}
	\frac{3}{2}\sum_{m=-\infty}^{\infty}\left[\frac{e^{ikR_m}}{kR_m}+\frac{1}{\bar{d}_0}\frac{\partial^2}{\partial(\bar{d}-m\bar{D})^2}\frac{e^{ikR_m}}{kR_m}\right]e^{i\beta mD}=
	\frac{3}{2}\sum_{n=-\infty}^{\infty}\frac{e^{-i\tilde{\beta}_{n}\bar{d}}}{\bar{D}}\frac{\pi}{i}\tilde{\beta}_{n}^{2}H_{0}^{(1)}\left[\bar{d}_0\sqrt{1-\tilde{\beta}_{n}^2}\right]
	\label{eq46}
	\end{equation}
Substituting the results into \Eq{eq45} we get
	\begin{equation}
	{M}_{n',q'}^{zz}=\frac{3}{2}\sum_{n=-\infty}^{\infty}\frac{e^{-i\tilde{\beta}_{n}\bar{d}}}{\bar{D}}\frac{\pi}{i}\left(\tilde{\beta}_{n}^2-1\right)H_{0}^{(1)}\left[\bar{d}_0\sqrt{1-\tilde{\beta}_{n}^2}\right]
	\label{eq47}
	\end{equation}

\subsection{Evaluation of ${M}_{n',q'}^{xz}$}\label{xz}
The value of the z component of $\bb{E}^L$ that the red chain creates at the origin due to the x components of the dipoles in the red chain 
	\begin{equation}
	\bb{E}_{z}^{L}(\bb{r}=(0,0,0))=\sum_{m=-\infty}^{\infty}G_{xz}(0,\bb{r}_m)p_{0x}e^{i\beta mD}
	\label{eq48}
	\end{equation}
where $p_{0x}e^{i\beta mD}$ are the z components of the dipole moments induced on the m'th particle in the red chain.
substituting $G_{xz}$ we may obtain an explicit form for ${M}_{n',q'}^{xz}$
	\begin{equation}
	\begin{split}
	{M}_{n',q'}^{xz}=\frac{3}{2}\sum_{m=-\infty}^{\infty}e^{ikR_m}\left[-\frac{\bar{d}_0\left(\bar{d}-m\bar{D}\right)}{(kR_m)^3}-3i\frac{\bar{d}_0\left(\bar{d}-m\bar{D}\right)}{(kR_m)^4}+3\frac{\bar{d}_0\left(\bar{d}-
	m\bar{D}\right)}{(kR_m)^5} \right]e^{i\beta mD}
	\label{eq49}
	\end{split}
	\end{equation}
\Eq{eq49} can also be presented as
	\begin{equation}
	{M}_{n',q'}^{xz}=\frac{3}{2}\sum_{m=-\infty}^{\infty}\left[\frac{\partial}{\partial\bar{d}_0}\frac{\partial}{\partial(\bar{d}-m\bar{D})}\frac{e^{ikR_m}}{kR_m}\right]e^{i\beta mD}
	\label{eq50}
	\end{equation}
Using previous results obtained we get
	\begin{equation}
	{M}_{n',q'}^{xz}=\frac{3}{2}\sum_{n=-\infty}^{\infty}\frac{e^{-i\tilde{\beta}_{n}\bar{d}}}{\bar{D}}\frac{\pi}{i}\left(i\tilde{\beta}_{n}\sqrt{1-\tilde{\beta}_{n}^2}\right)H_{1}^{(1)}\left[\bar{d}_0\sqrt{1-\tilde{\beta}_{n}^2}\right]
	\label{eq51}
	\end{equation}
\subsection{How many Hankel function terms shuold we sum?}
The Hankel function summations we have received are composed of terms of the general form
\begin{equation}
A_n=CQ(\tilde{\beta}_{n})H_{\nu}^{(1)}\left[\bar{d}_0\sqrt{1-\tilde{\beta}_{n}^2}\right]
\label{eq52}
\end{equation}
Where $Q$ may be a polynomial of order up to 2 or a combination of a polynomial with the square root function (Still, the maximal power of $\tilde{\beta}_{n}$ in $Q$ is 2), and $C$ is some complex constant. For the cases examined in sections ~\ref{xx} through ~\ref{xz} the "worst-case scenario" for the decay of the terms in the Hankel function series is when Q is a polynomial of order 2. for this case the series will have terms such as
	\begin{equation}
	A_n\propto\tilde{\beta}_{n}^2H_{\nu}^{(1)}\left[\bar{d}_0\sqrt{1-\tilde{\beta}_{n}^2}\right]
	\label{eq53}
	\end{equation}
or more explicitly
	\begin{equation}
	%\frac{2\pi n}{\bar{D}}-\frac{\beta}{k}
	A_n\propto\left(\frac{2\pi n}{\bar{D}}-\frac{\beta}{k}\right)^2H_{\nu}^{(1)}\left[\bar{d}_0\sqrt{1-\left(\frac{2\pi n}{\bar{D}}-\frac{\beta}
	{k}\right)^2}\right]
	\label{eq54}
	\end{equation}
%The largest absolute value for Hankel's function is obtained when $\left|{1-\left(\frac{2\pi n}{\bar{D}}-\frac{\beta}{k}\right)^2}\right|$ is minimal. this may occur for different n values, but for common choices of the geometrical parametes, we may assume that the maximal absolute value occurs for $n=0$. 
We may define the criteria for the number of terms to use for the summation as $n_0$ such that the tail of the summation $\sum_{n=n_0}^{\infty}A_n$ will be significantly smaller then a threshold constant C.
For large values of n we may use the large argument approximation for Hankel's function
\begin{equation}
H_{\nu}^{(1)}(z)\sim\sqrt{\frac{2}{\pi z}}e^{i\left(z-\frac{1}{2}\nu\pi-\frac{1}{4}\pi\right)}
\label{eq55}
\end{equation}
If we subsitute this approximation into the required sum we obtain
\begin{equation}
\left|\sum_{n=n_0}^{\infty}A_n\right|\leq\sum_{n=n_0}^{\infty}\left|A_n\right|\sim\sum_{n=n_0}^{\infty}\frac{\pi}{\bar{D}}\frac{n^{3/2}}{\sqrt{2\pi \frac{d_0}{D}}}e^{-2\frac{d_0}{D}\pi n}<C
\label{eq56}
\end{equation}
from \Eq{eq56} it is easy to see that the decay rate of the summation is strongly dependant on the ratio $\frac{d_0}{D}$. Evaluating the given summation may prove challenging, but if we don't mind giving a more strict criteria we may evaluate the needed $n_0$ from the following
\begin{equation}
\sum_{n=n_0}^{\infty}\frac{\pi}{\bar{D}}\frac{n^{3/2}}{\sqrt{2\pi \frac{d_0}{D}}}e^{-2\frac{d_0}{D}\pi n}<\sum_{n=n_0}^{\infty}\frac{\pi}{\bar{D}}\frac{n^{2}}{\sqrt{2\pi \frac{d_0}{D}}}e^{-2\frac{d_0}{D}\pi n}<C
\label{eq57}
\end{equation}
And the required value of $n_0$ may be extracted from the relation
\begin{equation}
\frac{\pi}{\bar{D}}\frac{1}{\sqrt{2\pi\frac{d_0}{D}}}\frac{e^{-2\pi\frac{d_0}{D}(n_0+1)}}{{\left(e^{-2\pi\frac{d_0}{D}}-1\right)^3}}\left[n_{0}^{2}e^{-4\pi\frac{d_0}{D}}+(-2n_{0}^{2}-2n_{0}+1)e^{-2\pi\frac{d_0}{D}}+(n_0+1)^2\right]<C
\label{eq58}
\end{equation}
\section{Example - Examination and verification of the magnetic model for particle rings}
\label{ParticleRingss}
\begin{figure}[htbp]
\centering
\includegraphics[scale=0.75]{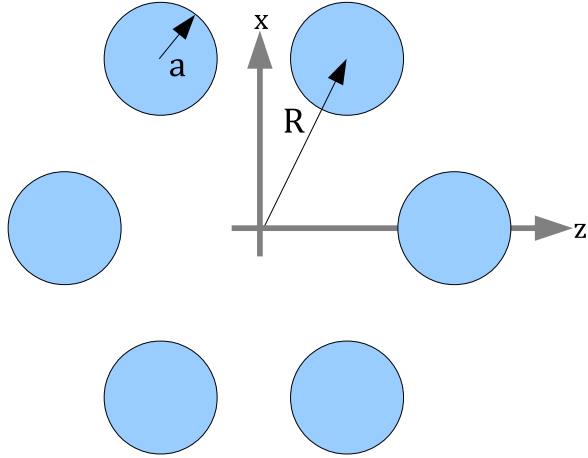}
\caption{Particle Ring}
\label{fig4}
\end{figure}

\begin{figure}[htbp]
\centering
\includegraphics[scale=0.75]{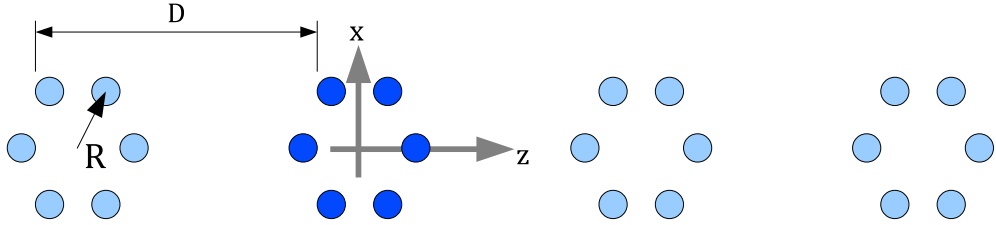}
\caption{A chain of particle rings}
\label{fig5}
\end{figure}

\begin{figure}[htbp]
\centering
%\hspace*{-0.5in}
\includegraphics[scale=0.8]{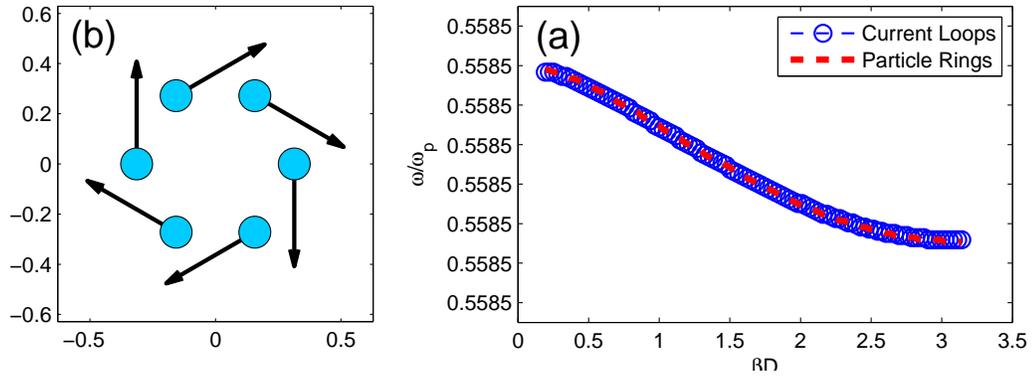}
\caption{(a) Dispersion curves as calculated with both methods described. (b) The dipole vectors as calculated for a solution on the dispersion curve}
\label{fig6}
\end{figure}

In \cite{EnghetaRings} a magnetic polarizability model for particle rings is developed. Each ring is composed of N spherical particles of radius $a$, positioned on the circumference of a ring with radius $R$ with uniform angular shift from each other. The electric dipoles induced on each particle in this case are in a direction tangent to the ring. Example for the setup for such a particle ring with 6 particles which we will use for the example is given in Fig.~\ref{fig4}. 
When exposed to y-polarized magnetic field ${H}_0\hat{y}$, an equivalent magnetic dipole moment will be induced on the particle ring that satisfies the relation $\alpha_{m}\bb{H}_0=m_0$ where the inverse magnetic polarizability is given by
\begin{equation}
\frac{4\pi}{k^3}\alpha_m^{-1}=\frac{8}{3N(kR)^2}\left[\frac{6\pi\epsilon_0}{k^3}\alpha_p^{-1}\right]
-\frac{1}{4N(kR)^5}\sum_{m=1}^{N-1}\frac{e^{2ikR\sin{m\frac{\pi}{N}}}}{sin^3\left(m\frac{\pi}{N}\right)}Q_m
\label{eq59}
\end{equation}
Where
\begin{equation}
\begin{split}
Q_m=\left\{3-3(kR)^2+\left(1+4(kR)^2\right)cos(2m\frac{\pi}{N})-\right.\\
\left.-kR\left(kRcos(4m\frac{\pi}{N}) +i\left[5sin(m\frac{\pi}{N})+sin(3m\frac{pi}{N})\right]\right)\right\}
\end{split}
\end{equation}
And $\alpha_p$ is the electric polarizability of the particles composing the ring. The dispersion curve for a chain of such rings may be calculated in two ways. The first, is by taking the equivalent magnetic polarizability of such a ring, and assuming the chain is a simple chain composed of magnetic particles that have the given polarizability. This way is very simple and streightforward. The second, by taking a particle-by-particle model of such a ring, using the formulation developed here previously. Ofcourse the second method contains many more degrees of freedom and the dispersion curve will most certainly contain many branches, but we expect one of these branches to be consistent with the model in \cite{EnghetaRings}, and moreover, the nullspace vector for that branch will exhibit tangent vectors, that represend the tangent dipoles. The chain setup is demonstated in Fig.~\ref{fig5}.
Dispersion curves for such chain of rings is given in Fig.~\ref{fig6}.  Its very clear that the 2 methods provide dispersion curves that overlap. calculating the nullspace vector for some selected solution from the dispersion curves shows dipole vectors tangent to the particle ring, which proves that this curve represents \emph{magnetic mode} that's propagating along the chain. the parameters used for the examined chain: $D=\frac{\lambda_p}{20},\;R= \frac{\lambda_p}{200},\;a=\frac{\lambda_p}{800}$.
\newpage

\end{document}